\documentclass{appolb}
\usepackage{epsfig}

\begin{document}
\title{Astrophysics of the Knee in the Cosmic Ray Energy Spectrum%
\thanks{Presented at the ISMD 2003, Krakow, Poland}%
}
\author{
A.~Haungs$^1$, \\
T.~Antoni$^2$,
W.D.~Apel$^1$, 
F.~Badea$^1$, 
K.~Bekk$^1$,
A.~Bercuci $^1$,
H.~Bl\"umer$^{1,2}$, 
H.~Bozdog$^1$, 
I.M.~Brancus$^3$,
C.~B\"uttner$^2$,
A.~Chilingarian$^4$,
K.~Daumiller$^2$,
P.~Doll$^1$,
R.~Engel$^1$,
J.~Engler$^1$,
F.~Fe{\ss}ler$^1$, 
H.J.~Gils$^1$,
R.~Glasstetter$^1$,\thanks{now at: Fachbereich Physik, 
Universit\"at Wuppertal, 42097
Wuppertal, Germany}, 
D.~Heck$^1$,
J.R.~H\"orandel$^2$, 
K.-H.~Kampert$^{1,2,\dag}$,
H.O.~Klages$^1$,
G.~Maier$^1$,
H.J.~Mathes$^1$, 
H.J.~Mayer$^1$, 
J.~Milke$^1$, 
M.~M\"uller$^1$,
R.~Obenland$^1$,
J.~Oehlschl\"ager$^1$,
S.~Ostapchenko$^2$,
M.~Petcu$^3$,
S.~Plewnia$^1$,
H.~Rebel$^1$,
A.~Risse$^5$,
M.~Risse$^1$,
M.~Roth$^2$,
G.~Schatz$^1$,
H.~Schieler$^1$,
J.~Scholz$^1$,
T.~Thouw$^1$,
H.~Ulrich$^1$,
J.~van~Buren$^1$,
A.~Vardanyan$^4$,
A.~Weindl$^1$,
J.~Wochele$^1$,
J.~Zabierowski$^5$,
S.~Zagromski$^1$\address{%
$^1$Institut\ f\"ur Kernphysik, Forschungszentrum 
Karlsruhe, Karlsruhe, Germany \\
$^2$Institut f\"ur Experimentelle Kernphysik,
Univ Karlsruhe, Karlsruhe, Germany \\
$^3$National Institute of Physics and Nuclear Engineering,
7690~Bucharest, Romania \\
$^4$Cosmic Ray Division, Yerevan Physics Institute, Yerevan~36, 
Armenia \\
$^5$Soltan Institute for Nuclear Studies, 90950~Lodz, 
Poland%
}
}
\maketitle
\begin{abstract}
\vspace*{-0.5cm}
A brief review is given on the astrophysics of cosmic rays in the 
PeV primary energy range, i.e.~the region of the knee.
\end{abstract}
  
\section{Introduction}
The all-particle energy spectrum of cosmic rays shows a 
distinctive feature at a few PeV, known as the knee, where the 
spectral index of a power-law dependence
changes from $-2.7$ to approximately $-3.1$. 
At that energy direct
measurements via balloon or satellite borne experiments 
are presently hardly possible due to the low flux. But indirect
measurements via the observation of extensive air showers (EAS)  
have been performed. 
Despite of more than 50 years of EAS measurements the origin of 
the knee is still unclarified, due to the difficult but 
inevitable disentanglement of the reconstruction of the energy 
and mass of the incoming primary particle with the necessary 
understanding of the air-shower development in the 
Earth's atmosphere.
This disentanglement remains an important experimental and 
theoretical challenge. 
A general introduction to the subject can be found in a recent 
review~\cite{rpp}. \\
To solve the puzzle of the knee the experimental access is the
reconstruction of energy spectra for individual elements 
(or mass groups), with an accompanying careful investigation of the 
hadronic interaction mechanisms governing the 
air-shower development. \\
In the present contribution a short overview on theoretical
ideas of explaining the knee is given. Additionally recent results 
from the advanced air-shower experiment KASCADE are discussed and 
compared to astrophysical model predictions. 
Furthermore the connection of the source of charged cosmic rays with 
measurements of TeV gamma-rays is outlined.  
Hadronic interaction models used for the simulation of the 
air-shower development are needed in all analyses of the data of 
air-shower experiments. Hence, tests of their validity is a crucial 
item of the data reconstruction. 
Details of such investigations are described in the
contributions of R.Engel et al.~\cite{ismd:ralph} and 
J.Milke et al.~\cite{ismd:jens} at this symposium.  

\section{Theoretical attempts for explaining the origin of the knee}
Theories on market about the origin of the knee can be grouped in 
three classes:
\begin{enumerate}
\item By acceleration: 
The knee energy is the maximum energy reached by 
acceleration of cosmic rays in our galaxy. This maximum energy is 
defined by the size and magnetic field strength
of the acceleration region and depends on the charge Z of the 
primary particles ($E_{max} \propto Z \times (L \times B)$). 
For example, Biermann et al.~\cite{biermann} proposed a scenario 
with a two component supernova acceleration, where 'normal' 
supernovae explode into the interstellar 
medium and accelerate mainly protons up to approximately 100 TeV and 
where more massive supernovae exploding into their own stellar wind  
accelerate also heavier particles up to $Z \cdot 10^{15}\,$eV.   
\item By diffusion: 
The idea is that the magnetic field retains the particles within the
Galaxy up to energies
at the knee ($E_{max} \approx Z \cdot 3 \cdot 10^{15}\,$eV). Particles of 
higher energy would start to escape from our galaxy. 
A detailed calculation has been performed by 
Candia et al.~\cite{candia}
starting with a constant source spectrum at 1 TeV (from results of 
direct measurements) and taking into account a regular plus an 
overlaying turbulent component of the galactic magnetic field. 
Fig.~\ref{candia:fig}
\begin{figure}
\begin{center}
\resizebox{0.8\textwidth}{!}{%
  \includegraphics{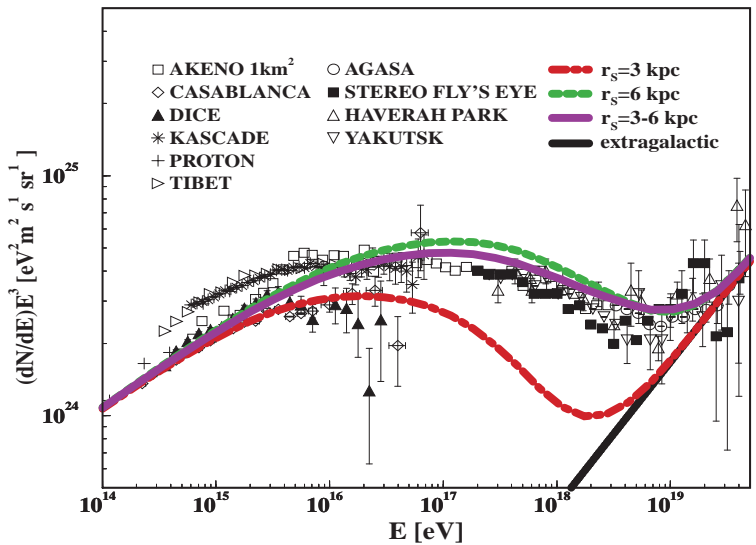}
}
\end{center}
\vspace*{-0.5cm}
\caption{Comparison of the measured cosmic-ray all-particle energy 
spectrum with predictions by a diffusion model 
(from~\cite{candia}).}
\label{candia:fig}   \vspace*{-0.3cm}   
\end{figure}
shows predictions for the all-particle energy spectrum of cosmic rays
for different assumptions of average distances of the sources. 
Predictions for large scale anisotropies by the diffusion are also
given in~\cite{candia}. 
\item By hadronic interaction features: 
Such models assume that a 
new channel of interaction is opening, either in 
the interstellar medium or inside our atmosphere. In this new 
channel a part of the primary energy is dissociated, and therefore 
unseen in the air-shower. 
This leads to the reconstruction of too small primary energies. 
Such a mechanism should start at energies proportional to the mass of 
the primaries, i.e.~the knee position should shift in proportion
to the atomic number 
and not to the charge of the primary particles. 
Possible scenarios are here the production of gravitons
in pp collisions~\cite{kazanas} or interactions of the primaries 
with heavy relic neutrinos~\cite{wigmans}. Most probably such
models will be confirmed or excluded by data from the LHC. 
\end{enumerate}
To distinguish between these mentioned theories, several burning 
questions have to be answered by experiments measuring air-showers 
in the PeV energy range:
\begin{itemize}
  \parskip0pt
  \itemsep0cm
\item The all-particle energy spectrum and possible fine structures: 
Is the knee sharp or smooth with exact power laws below and
above? At which energy is the position of the knee?
\item The chemical composition around the knee: 
Does it change with energy? Does the mean mass become heavier or 
lighter?
\item Energy spectra of single elements or mass groups: 
Do all primary mass groups show a knee feature? Do the spectra 
follow power laws above the knee? 
Do different primaries exhibit a knee at 
different energies? If yes, do the positions scale with mass or 
charge of the primary particles?
\item The proton spectrum: 
Where is the knee of the proton spectrum? Is there more than 
one knee?
\item Isotropy:
Are the cosmic rays distributed isotropically over the whole energy 
range of the knee? Or do large scale anisotropies occur? Are any 
point source visible?
\item Primary photons:
Are there high-energy gamma rays as primary particles? 
If yes, is their origin diffuse or do they originate from 
point sources?
\item Air-shower development: 
Is the air-shower development driven by the hadronic interactions
of high-energy particles well understood? 
\end{itemize} 
Due to the indirect nature of EAS measurements the latter point 
hampers a definite conclusion on most of the above mentioned 
questions. Here a co-operation between the accelerator and 
cosmic-ray physics communities is highly desired~\cite{needs}.

\section{Results of the KASCADE experiment}
\begin{figure}
\begin{center}
\resizebox{0.99\textwidth}{!}{%
  \includegraphics{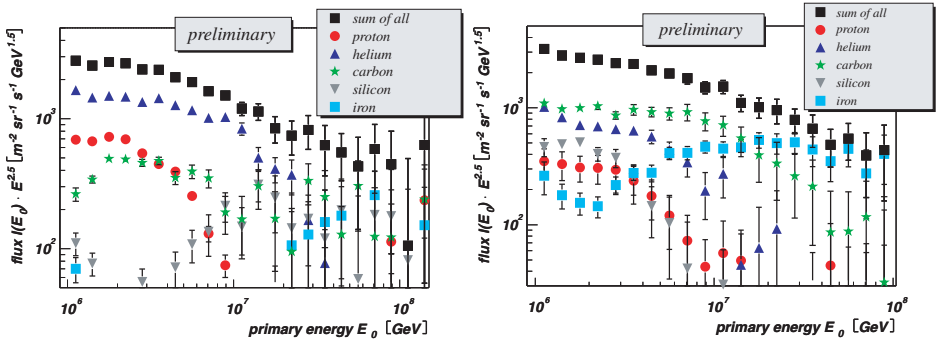}
}
\end{center}
\vspace*{-0.5cm}
\caption{Results of the unfolding procedure. Left panel: analyses 
based on QGSJet; right panel: based on SIBYLL.}
\label{spectra:fig}   \vspace*{-0.3cm}   
\end{figure}

KASCADE (KArlsruhe Shower Core and Array DEtector) is an 
air-shower experiment with a sophisticated detector set up
for detailed investigations 
of primary cosmic rays in the energy range of the knee. 
For the reconstruction of the energy and mass of the primary
particles and for the investigation of high-energy hadronic 
interactions, KASCADE~\cite{kas} follows the concept of a 
multi-detector set-up to provide as much as possible redundant 
information for each single air-shower event. 
The multidetector system allows 
to measure the total electron ($E_e>5\,$MeV) 
and muon numbers ($E_\mu>240\,$MeV)
of the shower separately using an array of 252 detector stations 
in a grid of $200 \times 200\,$m. Additionally muon densities at 
\begin{figure}
\begin{center}
\resizebox{0.80\textwidth}{!}{%
  \includegraphics{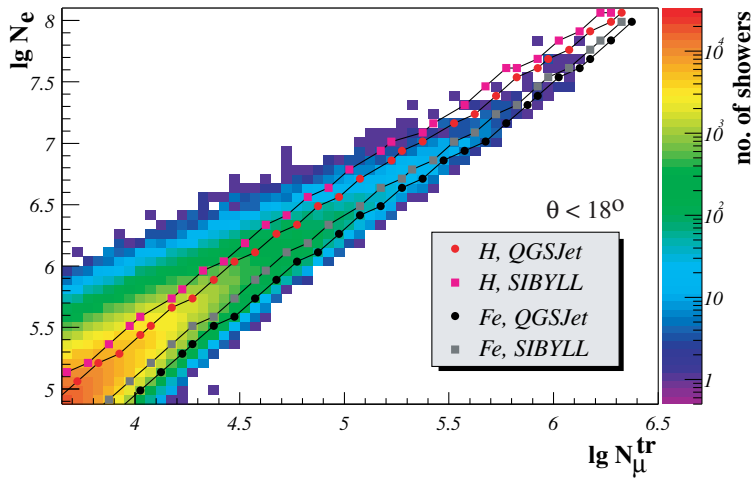}
}
\end{center}
\vspace*{-0.5cm}
\caption{Two dimensional electron ($N_e$) and muon 
($N_\mu^{\rm tr}$) number spectrum measured by the KASCADE array.
Lines display the most probable values for proton and iron simulations
of two different hadronic interaction models.}
\label{fig:1}   \vspace*{-0.3cm}   
\end{figure}
three further muon energy thresholds and the hadronic core of the 
shower by a $300\,$m$^2$ iron sampling calorimeter are measured. \\
The basic analysis of KASCADE to obtain the energy and mass of the
cosmic rays is a procedure of 
unfolding the two-dimensional electron-muon number spectrum 
(Fig.~\ref{fig:1}) into the energy spectra of five primary mass 
groups~\cite{kas:ulrich}. 
The problem can be considered as 
a system of coupled Fredholm integral equations of the form \\
{\small  $\frac{dJ}{d\,\lg N_e d\,\lg N_\mu^{tr}} = 
  \sum_A \int\limits_{-\infty}^{+\infty} \frac{d\,J_A}{d\,\lg E} 
  \cdot 
  p_A(\lg N_e\, , \,\lg N_\mu^{tr}\, \mid \, \lg E)
  \cdot 
  d\, \lg E $ } \\
where the probability $p_A$ is a further integral with a kernel 
function factorized into three parts, describing 
1) the shower fluctuations, 
i.e.~the distribution of electron and muon numbers for fixed 
primary energy and mass; 
2) the trigger efficiency of the experiment; 
3)the reconstruction probabilities, 
i.e.~the distributions of reconstructed $N_e$ and $N_\mu^{\rm tr}$
for given true numbers of electrons and muons.
The probabilities $p_A$ are obtained by extensive 
Monte Carlo simulations partly followed by a detailed 
detector simulation based on GEANT. 
The application of the unfolding procedure to the data 
is performed on basis of 
two different hadronic interaction models (QGSJet~\cite{qgs} and 
SIBYLL~\cite{sib}) as options embedded in CORSIKA~\cite{cors} for 
the reconstruction of the kernel functions.
Fig.~\ref{fig:1} displays the mean distributions of the generated showers in the
electron-muon number plane for both interaction models 
and for proton and iron primaries. 
Obviously the differences for both model
leads to differences of the results of the unfolding procedure,
as seen in Fig.~\ref{spectra:fig}. \\
It is worthwhile to note that despite of the large differences in the 
relative abundances of the primary mass groups, the all-particle 
energy spectrum and the fact that the knee is caused by the 
decreasing flux of 
light primaries, are very similar for both models. 
Results of further tests using different 
data sets, different unfolding methods, etc. show the same 
behavior~\cite{kas-roth}. \\ 
None of the present hadronic interaction models
can describe our multidimensional data consistently. 
Nevertheless there is, 
together with correlation information of the parameters used for 
model tests, no need to introduce a general new feature of 
the hadronic interactions, to account for our data. \\ 
The finding of the knee caused by light primaries is 
corroborated by results of an analysis of muon density 
measurements at KASCADE~\cite{kas-muon}, which were performed 
independently of the present reconstructions. 
With a much smaller influence of Monte Carlo simulations it 
could be shown, that data samples with enhanced light primaries 
show a knee feature, whereas samples with enhanced heavy primaries 
do not show so up to 10 PeV.\\
The air-shower events registered by KASCADE were additionally
analyzed in terms of large scale anisotropies and point source
signals. Within the statistical limits no deviation from 
a global isotropy of the arrival directions could be 
found~\cite{kas-maier}, but the statistical sensitivity of KASCADE
is not high enough to confirm or disproof the predictions 
of Candia et al.~\cite{candia}.   \\
There is also no positive evidence from KASCADE 
for primary photons in the 
PeV energy range~\cite{kas-fessler}, but the upper limits on 
the galactic diffuse gamma ray flux could be noticeably improved 
as compared to other experiments. \\

The astrophysical results so far available
from KASCADE can be summarized as: 
The knee in the PeV range is caused by the decrease of the 
flux of light particles, and heavier (A$>$20) primary particles 
exhibit no knee up to 10 PeV. This leads to an increase of the mean 
mass of cosmic rays when passing the knee region in energy.
Due to the present uncertainties of the hadronic interactions in the
atmosphere (e.g.~the high-energy extrapolations, 
the diffraction cross-section and multiplicity parametrisations
of the models) it is presently not possible to give more refined 
quantitative results.

\section{The position of the proton knee}
The most abundant particles in the TeV range of cosmic rays are 
protons. Hence there is a high interest in the proton spectrum
in the range of the crossover from direct to air-shower measurements,
especially in view of various theories predicting
different positions of the proton knee, depending on the magnetic 
field strength and size of the acceleration or diffusion region. \\
Some of the direct measurements favor a proton knee at 10
TeV~\cite{zatsepin},
recent results from the Tibet air-shower experiment
(sensitive to lower energies than KASCADE due to the larger 
observation height of $4300\,$m) claim
the proton knee at around 500 TeV~\cite{tibet}, 
whereas KASCADE observes a change of the spectrum at 
3-5 PeV~\cite{kas:ulrich}.
Measurements of a  more detailed structure of the proton spectrum
therefore are of astrophysical importance. But due
to different measurement techniques they are difficult to compare.
To establish a more comprehensive picture of the proton spectrum 
from 1 TeV to 100 PeV is one of the main tasks of future cosmic 
ray experiments.
 
\section{Charged cosmic rays and TeV gamma ray astronomy}
Supernova shock acceleration is believed to be the main source of the 
cosmic rays in our galaxy. But positive evidence 
for proton acceleration at these kind of objects is still missing. 
\begin{figure}
\begin{center}
\resizebox{0.8\textwidth}{!}{%
  \includegraphics{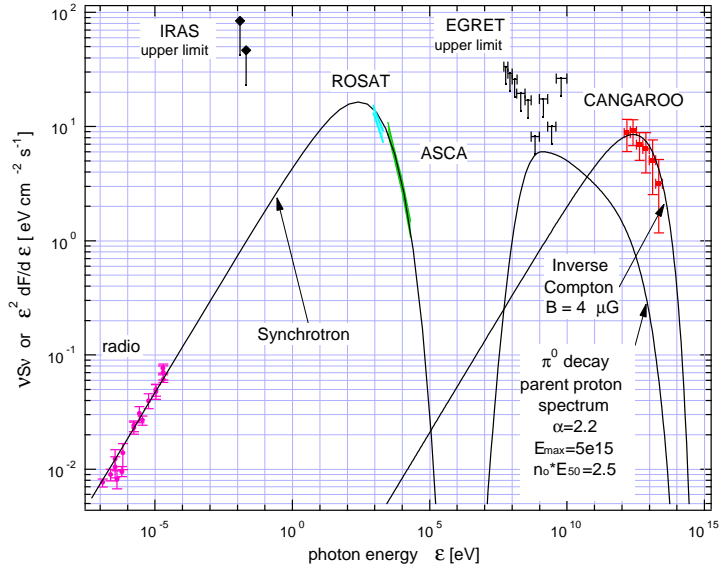}
}
\end{center}
\vspace*{-0.5cm}
\caption{Multi-band spectrum of energy fluxes from SN1006 by different
experiments. Solid lines are fits to models of inverse Compton gamma
production and pion decay production, respectively
(from~\cite{cangaroo}).}
\label{gamma:fig}   \vspace*{-0.3cm}   
\end{figure}
If protons are accelerated to PeV energies some of them should 
interact with the surrounding matter producing $\pi^0$ which 
decay into high-energy gammas in the TeV range. \\
In the last decade there has been a lot of progress in TeV-gamma ray
astronomy~\cite{ong}, and indeed supernova remnants were found 
as sources of TeV-gammas. But by comparing the spectra with 
measurements at lower frequency bands, the TeV gamma ray fluxes can 
be consistently explained by a
'self-synchrotron--inverse-compton' model, i.e.~by electron 
acceleration without any contribution from pion 
decays~\cite{cangaroo} (Fig.~\ref{gamma:fig}). 
Just recently Berezhko and V\"olk~\cite{volk} 
suggested a theoretical picture 
of a more efficient proton acceleration inside the sources which 
can explain the measured spectra also with proton acceleration 
and a smaller contribution of inverse Compton produced gammas. \\  
Further measurements of this kind provide interesting aspects in 
understanding the source of charged cosmic rays. 

\section{Future prospects}
The investigations of charged cosmic rays around the knee will be 
continued and improved by the KASCADE-Grande 
experiment~\cite{kas:kg,kas:kg1}, which
is an extension of KASCADE to measure air-showers up to primary
energies of 1 EeV. KASCADE-Grande is also an multi-detector setup,
and therefore correlation analyses are able to check the validity 
of hadronic interaction models. This provides complementary
information to the LHC measurements due to the extreme forward
physics of the air-shower development. \\
Concerning direct measurements new technical issues may allow
to fly larger detectors during longer flights, which would increase
the statistical accuracy at higher energies.
Also new TeV-gamma ray experiments with higher sensitivity 
presently under construction will
contribute to solve the puzzle of source, acceleration, and transport
of high energy cosmic rays, and in particular the origin of the knee.

\end{document}